\documentclass[aps,pra,twocolumn,nofootinbib]{revtex4-2}
\usepackage[colorlinks]{hyperref}
\usepackage[utf8]{inputenc}
\usepackage{graphicx}
\usepackage{multirow}
\usepackage{amssymb}
\usepackage{amsmath}
\usepackage{microtype}
\usepackage{bm}
\usepackage{bbm}
\usepackage{enumitem}
\usepackage{bbold}

\begin{document}
\newcommand{\half}{\frac12}
\newcommand{\vare}{\varepsilon}
\newcommand{\eps}{\epsilon}
\newcommand{\pr}{^{\prime}}
\newcommand{\ppr}{^{\prime\prime}}
\newcommand{\pp}{{p^{\prime}}}
\newcommand{\ppp}{{p^{\prime\prime}}}
\newcommand{\hp}{\hat{\bfp}}
\newcommand{\hr}{\hat{\bfr}}
\newcommand{\hk}{\hat{\bfk}}
\newcommand{\hx}{\hat{\bfx}}
\newcommand{\hpp}{\hat{\bfpp}}
\newcommand{\hq}{\hat{\bfq}}
\newcommand{\rqq}{{\rm q}}
\newcommand{\bfk}{{\bm{k}}}
\newcommand{\bfp}{{\bm{p}}}
\newcommand{\bfq}{{\bm{q}}}
\newcommand{\bfr}{{\bm{r}}}
\newcommand{\bfx}{{\bm{x}}}
\newcommand{\bfy}{{\bm{y}}}
\newcommand{\bfz}{{\bm{z}}}
\newcommand{\bfpp}{{\bm{\pp}}}
\newcommand{\bfppp}{{\bm{\ppp}}}
\newcommand{\balpha}{\bm{\alpha}}
\newcommand{\bvare}{\bm{\vare}}
\newcommand{\bgamma}{\bm{\gamma}}
\newcommand{\bGamma}{\bm{\Gamma}}
\newcommand{\bLambda}{\bm{\Lambda}}
\newcommand{\bmu}{\bm{\mu}}
\newcommand{\bnabla}{\bm{\nabla}}
\newcommand{\bvarrho}{\bm{\varrho}}
\newcommand{\bsigma}{\bm{\sigma}}
\newcommand{\bTheta}{\bm{\Theta}}
\newcommand{\bphi}{\bm{\phi}}
\newcommand{\bomega}{\bm{\omega}}
\newcommand{\intzo}{\int_0^1}
\newcommand{\intinf}{\int^{\infty}_{-\infty}}
\newcommand{\lbr}{\langle}
\newcommand{\rbr}{\rangle}
\newcommand{\ThreeJ}[6]{
        \left(
        \begin{array}{ccc}
        #1  & #2  & #3 \\
        #4  & #5  & #6 \\
        \end{array}
        \right)
        }
\newcommand{\SixJ}[6]{
        \left\{
        \begin{array}{ccc}
        #1  & #2  & #3 \\
        #4  & #5  & #6 \\
        \end{array}
        \right\}
        }
\newcommand{\NineJ}[9]{
        \left\{
        \begin{array}{ccc}
        #1  & #2  & #3 \\
        #4  & #5  & #6 \\
        #7  & #8  & #9 \\
        \end{array}
        \right\}
        }
\newcommand{\Vector}[2]{
        \left(
        \begin{array}{c}
        #1     \\
        #2     \\
        \end{array}
        \right)
        }

\newcommand{\Dmatrix}[4]{
        \left(
        \begin{array}{cc}
        #1  & #2   \\
        #3  & #4   \\
        \end{array}
        \right)
        }
\newcommand{\Dcase}[4]{
        \left\{
        \begin{array}{cl}
        #1  & #2   \\
        #3  & #4   \\
        \end{array}
        \right.
        }
\newcommand{\cross}[1]{#1\!\!\!/}

\newcommand{\Za}{{Z \alpha}}
\newcommand{\im}{{ i}}

\title{Calculation of isotope shifts and King plot nonlinearities in Ca$^+$}
	
\author{A.~V.~Viatkina$^{1,2}$}
\author{V.~A.~Yerokhin$^{3}$}
\author{A.~Surzhykov$^{1,2}$}

\affiliation{$^1$Technische Universität Braunschweig, 38106 Braunschweig, Germany}
\affiliation{$^2$Physikalisch-Technische Bundesanstalt, 38116 Braunschweig, Germany}
\affiliation{$^3$Max-Planck-Institut für Kernphysik, Saupfercheckweg 1, 69117 Heidelberg, Germany}
	
\begin{abstract}
Many-body perturbation theory is implemented in order to calculate the isotope shifts of $4s$, $4p_{1/2}$, $4p_{3/2}$, $3d_{3/2}$, and $3d_{5/2}$ energy levels of Ca$^+$, for even isotopes $A=$40, 42, 44, 46, 48. The results are presented for mass shift and field shift, as well as for higher-order field shifts, quadratic mass shift, nuclear polarization correction, and the cross term between field and mass shifts. Additionally, we examine King-plot nonlinearities introduced by higher-order isotope-shift corrections to the combinations of $3d_{3/2}\rightarrow 4s$, $3d_{5/2}\rightarrow 4s$, and $4p_{1/2}\rightarrow 4s$ transitions. For these transitions, second-order mass shift and nuclear polarization correction are identified as the dominant sources of King plot nonlinearity.
\end{abstract}
	
\maketitle
	
\section*{Introduction}
High-precision atomic spectroscopy is currently in the spotlight of experimental research, particularly as a means of testing physics beyond standard model \cite{safronova_search_2018,berengut_probing_2018,counts_evidence_2020}. The recent advances in experimental technique allows one to determine atomic transitions with high accuracy which in certain cases reaches $0.1$~Hz~\cite{schmidt_spectroscopy_2005,wineland_quantum_2002,king_optical_2022,ludlow_optical_2015,picque_frequency_2019,kozlov_highly_2018}. One of the precision-spectroscopy methods aimed at detecting new interactions is King plot analysis; the basic idea was proposed in Ref.~\cite{delaunay_probing_2017,berengut_probing_2018} and later refined in Refs.~\cite{berengut_generalized_2020,counts_evidence_2020,drake_king_2021}. Roughly speaking, King plot is a plot constructed from normalized isotope shifts of atomic transitions which are measured in a series of isotopes of a given element; such plots can be proven to be linear to a high precision. However, it has been shown that an interaction between the electron cloud and the nucleus mediated by a hypothetical light boson would distort this linearity~\cite{berengut_probing_2018}. Hence, if a nonlinearity in a King plot is detected in experiment, it could be attributed to a new boson, given that any other explanation is ruled out \cite{berengut_probing_2018}. On the other hand, as one may expect, there are numerous sources of King-plot nonlinearities found already in the framework of standard model.

Recently, a significant King-plot nonlinearity has been discovered in Yb and Yb$^+$ transitions \cite{counts_evidence_2020,hur_evidence_2022,figueroa_precision_2022} and the efforts of its interpretation are still ongoing, see e.g.~Refs.~\cite{hur_evidence_2022, allehabi_nuclear_2021}. In contrast, King plots in a succession of Ca$^+$ isotopes were found to be linear within experimental uncertainties \cite{knollmann_part-per-billion_2019,knollmann_erratum_2023,solaro_improved_2020}. Isotope shifts were measured for $4s\ ^{2}S_{1/2}\rightarrow 4p\ ^{2}P_{1/2}$ and $3d\ ^2D_{3/2}\rightarrow 4p\ ^2P_{1/2}$ transitions with accuracy below 100~kHz~\cite{gebert_precision_2015,muller_collinear_2020}, for $4s\ ^{2}S_{1/2}\rightarrow 3d\ ^2D_{5/2}$ with accuracy below 10~Hz~\cite{knollmann_part-per-billion_2019}, and for the interval $3d\ ^2D_{3/2}- 3d\ ^2D_{5/2}$ with $\sim 20$~Hz accuracy \cite{solaro_improved_2020}.
One can expect that in the future the measurement accuracy reaches 1~Hz; in such case, the King-plot linearity might no longer hold and a careful theoretical examination of the origins of nonlinearity will be needed. In this contribution, therefore, we aim to lay ground for an analysis of Ca$^+$ King plot nonlinearity.

It should be noted that, as a system intended for isotope-shift spectroscopy, Ca$^+$ has several advantages. First, calcium has five even-even stable ($A=40,42,44,46$) or long-lived ($A=48$) isotopes, which is a welcome fact for building a King plot. Isotopes with zero nuclear spin are preferred for King-plot analysis, because their spectra lack hyperfine splitting. Second, the electronic structure and atomic transitions of Ca$^+$ are theoretically well-understood, given its alkali-like electronic configuration. Third, Ca$^+$ is a convenient object for experimental study: as was mentioned above, a series of experiments has already been conducted to determine several of its atomic transitions with high accuracy \cite{gebert_precision_2015,knollmann_part-per-billion_2019,knollmann_erratum_2023,muller_collinear_2020,solaro_improved_2020}, with a realistic possibility for future improvement.

In the present work, we calculate the main contributions to isotope shift in Ca$^+$ ions and evaluate their impact on King-plot linearity. The relativistic units ($\hbar=c=m=1$) are used throughout this paper, unless explicitly specified. The paper is structured as follows. In Sec.~\ref{sec:IS_overall} we introduce theoretical origins of isotope shift terms: we discuss mass shift (Sec.~\ref{sec:MS}), field shift (Sec.~\ref{sec:FS}),  nuclear polarization, and field- and mass-shift cross term (Sec.~\ref{sec:AT}). The total isotope shift---the sum of all considered contributions---is presented in Sec.~\ref{sec:all_contrib}. In Sec.~\ref{sec:mbpt} the specific methods for many-body calculations are laid out, which we subsequently use in Sec.~\ref{sec:NUM} to evaluate each of the isotope shift terms. Finally, we discuss Ca$^+$ King plots and their nonlinearity in Sec.~\ref{sec:KP} and summarize our results in Sec.~\ref{sec:conclusion}.



\section{Isotope shift: theory}\label{sec:IS_overall}
In the first-order approximation, the isotope shift of an energy level in the $i$-th isotope with respect to a reference isotope $a$ can be written as
\begin{equation}
    \Delta E_{ia}=K \left(\frac{m}{M_i}-\frac{m}{M_a}\right)+F \left(R^2_i-R^2_a\right),\label{eq:IS_1st}
\end{equation}
where $m$ is electron mass and $M_j$ denote nuclear masses ($j=a, i$). Here, we introduce a dimensionless nuclear charge radius
\begin{equation}
R_j^2=\langle r^2\rangle_j/\lambdabar^2_C\ ,\label{eq:R_def}
\end{equation}
which is a mean square nuclear charge radius $\langle r^2\rangle_j$ divided by (the square of) reduced Compton wavelength of an electron, $\lambdabar_C$. The first term in Eq.~\eqref{eq:IS_1st} corresponds to \textit{mass shift} (MS) and the second term to \textit{field shift} (FS); the former is the consequence of the change in nuclear mass and the latter of the change in nuclear charge distribution between two isotopes. Note that in the first-order approximation \eqref{eq:IS_1st} the electronic structure constants $K$ and $F$ do not depend on the nuclear parameters of the $i$-th isotope, while they may still implicitly include the parameters of the reference isotope $a$. Effectively, Eq.~\eqref{eq:IS_1st} assumes that electronic wavefunctions do not yet `notice' the change in nuclear mass and shape between isotopes.

In the present work, we investigate the effects beyond the first-order approximation \eqref{eq:IS_1st}. We take into account that both electronic `constants' $K$ and $F$ depend on the isotope $i$ in question and hence the energy shift can be written as
\begin{equation}
    \Delta E_{ia}=K_{ia} \left(\frac{m}{M_i}-\frac{m}{M_a}\right)+F_{ia} \left(R^2_i-R^2_a\right)\ . \label{eq:IS_full}
\end{equation}
Below, we will examine the isotope dependence of $K_{ia}$ and $F_{ia}$, treating mass shift and field shift separately. Moreover, we will consider further contributions to isotope shifts which do not, strictly speaking, belong either to mass or field shift: nuclear polarization correction and cross term between field and mass shifts. At the end, we will show how to extend Eq.~\eqref{eq:IS_full} to take those additional effects into account. Note that, in this work, we will use $^{40}$Ca isotope as our reference isotope $a$ ($A_a=40$).

\subsection{Mass shift}\label{sec:MS}
Isotope mass shift arises from the difference in nuclear recoil effect between two isotopes. For light atomic systems it is sufficient to describe the nuclear recoil by means of nonrelativistic operators \cite{tupitsyn_relativistic_2003}. Thus, let us write the Schr\"{o}dinger Hamiltonian of the atom:
\begin{equation}
    H=\frac{\vec{P}^2}{2M}+\sum_k\frac{\vec{p}_k^{\ 2}}{2m} + V_C\ .\label{eq:H_nonrel_0}
\end{equation}
Here $\vec{P}$ refers to the nuclear momentum, $\vec{p}_k$ to the momentum of the $k$-th electron. By $V_C$ we denote the Coulomb potential
\begin{equation}
    V_C=-\sum_k\frac{Z\alpha}{|\vec{r}_0-\vec{r}_k|}+\sum_{k<l}\frac{\alpha}{|\vec{r}_k-\vec{r}_l|}\ ,
\end{equation}
where $\alpha$ is the fine-structure constant and $\vec{r}_0$ and $\vec{r}_k$ are the position vectors of the nucleus and the $k$-th electron, respectively. Choosing the center-of-mass reference frame, we obtain $\vec{P} = -\sum_k \vec{p}_k$ and Eq.~\eqref{eq:H_nonrel_0} becomes \cite{johnson_atomic_2007}
\begin{equation}
    H=\sum_k\frac{\vec{p}_k^{\ 2}}{2m_r}+V_C+\frac{1}{M}V_\mathrm{SMS}\ ,\label{eq:H_nonrel_1}
\end{equation}
where $m_r=mM/(m+M)$ is the reduced mass. In this equation, the first term is the normal mass shift (NMS) operator, whose effect can be observed already in hydrogenlike systems, while the last term $V_\mathrm{SMS}$ is the many-electron specific mass shift (SMS) operator
\begin{equation}
    V_\mathrm{SMS}=\sum_{k<l}\vec{p}_k\cdot\vec{p}_l\ .
\end{equation}
To further evaluate \eqref{eq:H_nonrel_1}, we introduce $\mu=m_r/m$ and shift the variables in Eq.~\eqref{eq:H_nonrel_1} as $\vec{r}\rightarrow \mu^{-1} \vec{r}\,$ and, therefore, $\vec{p}\rightarrow \mu\vec{p}$. Thereby we obtain
\begin{align}
    H&=\mu\left[ \sum_k\frac{\vec{p}_k^{\ 2}}{2m}+V_C+\frac{\mu}{M}V_\mathrm{SMS} \right]\nonumber\\
    &\equiv \mu \left[ H_0 + \frac{\mu}{M}V_\mathrm{SMS} \right]\ ,
\end{align}
where $H_0$ is the nonrelativistic atomic Hamiltonian in the infinite nuclear mass limit. If we denote the eigenfunctions of $H_0$ as $\psi_0$, its eigenvalues as $E_0$, and expand the eigenvalues of $H$ in the powers of $m/M$, we obtain the first- and second-order nuclear recoil (or mass-shift) corrections to an electronic energy level:
\begin{align}
    \delta E_\mathrm{MS}^{(1)}&= \frac{m}{M}\left(-E_0+K^{(1)}_\mathrm{SMS}\right)\ ,\\
    \delta E_\mathrm{MS}^{(2)}&= \left(\frac{m}{M}\right)^2\left(E_0-K^{(1)}_\mathrm{SMS}+K^{(2)}_\mathrm{SMS}\right)\ .
\end{align}
Here, the first- and second-order specific mass shift constants are given by
\begin{align}
    K^{(1)}_\mathrm{SMS}&=\langle \psi_0|V_\mathrm{SMS}|\psi_0\rangle\ ,\\
    K^{(2)}_\mathrm{SMS}&={\sum_{n\neq 0}}\frac{|\langle \psi_0|V_\mathrm{SMS}|\psi_n\rangle|^2}{E_0-E_n}\ .  
\end{align}
By defining the MS constants of the first and second order
\begin{align}
    K^{(1)}&\equiv -E_0+K^{(1)}_\mathrm{SMS}\ ,\label{eq:MS_K1}\\
    K^{(2)}&\equiv E_0-K^{(1)}_\mathrm{SMS}+K^{(2)}_\mathrm{SMS}\ ,\label{eq:MS_K2}
\end{align}
we obtain an expression for the mass shift in an isotope $j$ relative to a hypothetical infinite-mass isotope:
\begin{equation}
    \Delta E_{\mathrm{MS}, j}=K^{(1)}\frac{m}{M_j}+K^{(2)}\frac{m^2}{M^2_j}\ .
\end{equation}
Accordingly, the mass shift of an isotope $i$ with respect to the reference isotope $a$ is
\begin{align}
    \Delta &E_{\mathrm{MS}, ia}=\nonumber\\
    &K^{(1)}\left(\frac{m}{M_i}-\frac{m}{M_a}\right)+K^{(2)}\left(\frac{m^2}{M_i^2}-\frac{m^2}{M_a^2}\right)\ .\label{eq:MS_isolated}
\end{align}
It should be noted that when calculating $K^{(1)}$ and $K^{(2)}$ in Sec.~\ref{sec:MS_num} and presenting the results in Table~\ref{tab:const} below, we use the experimental values of binding energies as our $E_0$. Strictly speaking, experimental energies $E$ are not equal to the eigenvalues $E_0$ of the infinite-mass-isotope Hamiltonian; however, the difference $|E-E_0|$ is negligible for our purposes, since the uncertainties of our many-body atomic calculations are much larger.

\subsection{Field shift}\label{sec:FS}
The nuclei of two isotopes differ not only in mass, but also in the parameters of the nuclear charge distribution. Electronic energy shifts which result from the difference in nuclear charge distribution---or, more precisely, from the difference in nuclear potential---are called field shifts (FS). For the purposes of the present investigation, it is sufficient to assume that all isotopes have the same shape of the nuclear charge distribution and differ only by the values of the charge radii. 
In this approximation, the electrostatic potential $V$ of a nucleus depends solely on the nuclear charge radius, $V=V(R)$; its total isotopic variation would be $\delta V_{ia}=V(R_i)-V(R_a)$. In the second-order perturbation theory, the field shift is given by
\begin{equation}
    \Delta E_{\mathrm{FS},ia}= \langle\psi_a|\delta V_{ia}|\psi_a\rangle +\sum_{n\neq a}\frac{|\langle\psi_a|\delta V_{ia}|\psi_n\rangle|^2}{E_a-E_n}\ .\label{eq:FS_2ndorder_initial}
\end{equation}
Here $\psi_a$ is the electronic wave function and $E_a$ the electronic energy level in the reference isotope~$a$.

Let $\delta R_{ia}^2=R_i^2-R_a^2$ be the difference between (squares of) dimensionless radii. In Eq.~\eqref{eq:FS_2ndorder_initial}, we would like to isolate the dominant first-order field shift $F^{(1)}\delta R_{ia}^2$ [see Eq.~\eqref{eq:IS_1st}] from other field-shift contributions. To do this, we introduce the standard field-shift operator $V_\mathrm{FS}\equiv\partial V(R)/\partial\left(R^2\right)$ and rewrite $\delta V_{ia}$ as
\begin{equation}
    \delta V_{ia}=V_\mathrm{FS}\,\delta R_{ia}^2+\left(\frac{\delta V_{ia}}{\delta R_{ia}^2}-V_\mathrm{FS}\right)\delta R_{ia}^2\ .\label{eq:FS_artificial}
\end{equation}
Now we can express the first term in Eq.~\eqref{eq:FS_2ndorder_initial} as follows:
\begin{equation}
\langle\psi_a|\delta V_{ia}|\psi_a\rangle = F^{(1)}(R_a)\delta R_{ia}^2+\delta_{R_i}F^{(1)}(R_a) \delta R_{ia}^2\ .
\end{equation}
Here, the first coefficient
\begin{equation}
    F^{(1)}(R_a)=\left<\psi_a\left|V_\mathrm{FS}\right|\psi_a\right>,\label{eq:F1_def}
\end{equation}
is the standard FS constant in Eq.~\eqref{eq:IS_1st}. The remaining part of the field shift is the higher-order correction
\begin{equation}
 \delta_{R_i}F^{(1)}(R_a)=\left<\psi_a\left|\frac{\delta V_{ia}}{\delta R_{ia}^2}-V_\mathrm{FS}\right|\psi_a\right>.\label{eq:dF1_ho}
\end{equation}

Since the second term of Eq.~\eqref{eq:FS_2ndorder_initial} is a small correction, one can replace $\delta V_{ia}$ by $V_{FS}$ in it. Hence, we introduce the second-order field-shift electronic constant
\begin{equation}
    F^{(2)}(R_a)=\sum_{n\neq a}\frac{|\langle\psi_a|V_\mathrm{FS}|\psi_n\rangle|^2}{E_a-E_n}\ .
\end{equation}
Finally, we can express the resulting field shift as
\begin{align}
   \Delta E_{\mathrm{FS},ia} =
    \left[F^{(1)}(R_a)+\delta_{R_i}F^{(1)}(R_a)\right] & \delta R_{ia}^2 \nonumber\\ + F^{(2)}(R_a)&\left(\delta R_{ia}^2\right)^2.
\end{align}

Note that the residual potential $\left(\delta V_{ia}/\delta R^2_{ia}-V_\mathrm{FS}\right)$ is localized in the nuclear region similarly to the dominant field-shift potential $V_\mathrm{FS}$. Hence, it makes sense to represent the higher-order correction $\delta_{R_i}F^{(1)}(R_a)$ as a factor $f_\mathrm{ho}\left(R_a,R_i\right)$ multiplied by the field-shift constant:
\begin{equation}
    \delta_{R_i}F^{(1)}\left(R_a\right)=-F^{(1)}\left(R_a\right)f_\mathrm{ho}\left(R_a,R_i\right)\cdot 10^{-3}\ .\label{eq:dF_ho}
\end{equation}
This form is convenient for presenting the results of our numerical calculation. We will report the numerical results for FS constants $F^{(1)}\left(R_a\right)$, $\delta_{R_i}F^{(1)}\left(R_a\right)$, and $F^{(2)}\left(R_a\right)$ in Sec.~\ref{sec:FS_num}.

\subsection{Additional terms}\label{sec:AT}
\subsubsection{Nuclear polarization}
Not only the nuclear shape, but also the disposition of a nucleus to be polarized by an electric field changes from isotope to isotope. Nuclear polarization (np) manifests itself in a correction to electronic energy levels; varying nuclear polarization between isotopes thus results in a contribution to isotope shift. Let $\zeta$ be an atomic state and $E_\zeta$ its energy; the np-correction to $E_\zeta$ would be \cite{plunien_nuclear_1991, nefiodov_nuclear_1996}
\begin{align}
    \Delta E_\mathrm{np}&= \nonumber\\
    -\alpha &\sum_{LM}B(EL)\sum_n\frac{|\left<\zeta|\mathcal{F}_LY_{LM}|n\right>|^2}{E_n-E_\zeta+\mathrm{sgn}(E_n)\omega_L}\ ,\label{eq:dE_np}
\end{align}
where $n$ runs over the complete spectrum of electronic states including positive and negative continuum, $L$ denotes the multipolarity of nuclear excitations, $B(EL)=B(EL;L\rightarrow 0)$ are the reduced probabilities of nuclear transitions from the excited (``$L$'') to the ground state (``$0$''), $\omega_L$ are the nuclear excitation energies with respect to the ground state, $Y_{LM}$ the spherical harmonics, and $\mathcal{F}_L$ are characteristic radial functions in the sharp-edge-nucleus approximation \cite{nefiodov_nuclear_1996}:
\begin{equation}
    \mathcal{F}_{L=0}\left(r\right)=\frac{5\sqrt{\pi}}{2r_0^3}\left[1-\frac{r^2}{r^2_0}  \right]\theta\left(r_0-r\right)\ ,
\end{equation}
\begin{align}
    \mathcal{F}_{L>0}\left(r\right)&=\frac{4\pi}{(2L+1)r_0^L}\left[
    \frac{r^L}{r_0^{L+1}}\theta\left( r_0-r\right)\right.\nonumber\\
    &+\left.\frac{r_0^L}{r^{L+1}}\theta\left(r-r_0\right)
    \right]\ ,
\end{align}
where $r_0=\sqrt{\langle r\rangle^2}$ is the radius of the nuclear sphere and $\theta$ the step function.

The main contributions to the sum in Eq.~\eqref{eq:dE_np} arise from two kinds of transitions: giant resonances and lowest-lying rotational transitions. The former are dominant because of their large transition strengths $B(EL)$, while the latter are enhanced due to the small denominator containing the transition frequency $\omega_L$. Accordingly, the dominant low-lying-level contribution to $\Delta E_\mathrm{np}$ in the vast majority of even-even nuclei comes from the electric quadrupole transition from the first rotational level to the ground state, $2^+\rightarrow 0^+$.

To facilitate numerical calculations in many-electron systems, we introduce a nuclear polarization potential $V_\mathrm{np}$ defined by its matrix elements between single-electron atomic states:
\begin{align}
    &\langle \zeta|V_\mathrm{np}|\xi\rangle=\nonumber\\
    -&\alpha \sum_{LM}B(EL)\sum_n\frac{\left<\zeta|F_LY_{LM}|n\right>\left<n|F_LY_{LM}|\xi\right>}{E_n-m+\mathrm{sgn}(E_n)\omega_L}\ .\label{eq:Vnp-Def}
\end{align}
Here we replaced $E_\zeta$ with electron mass, thereby discarding the (very weak) dependence of $V_\mathrm{np}$ on the binding energy of the state $\zeta$.

Similarly to the field-shift operator $V_\mathrm{FS}$, the operator $V_\mathrm{np}$ is mainly localized in the nuclear region. Hence it is convenient to present the np correction to a given energy level in Ca$^+$ as
\begin{equation}
    \Delta E_\mathrm{np}\equiv -g_{\mathrm{np},j}\, R_j^2\, F^{(1)}\left(R_a\right) \cdot 10^{-3}\ ,\label{eq:gNPone}
\end{equation}
where $R_j$ is the dimensionless nuclear charge radius \eqref{eq:R_def}, $F^{(1)}$ is defined in Eq.~\eqref{eq:F1_def}, and $g_{\mathrm{np},j}$ is a np coefficient which depends both on the isotope and on the electronic state in question. Accordingly, the \textit{isotope shift} due to nuclear polarization in a given calcium isotope $i$ with regards to the isotope $a$ would be:
\begin{equation}
    \Delta E_{\mathrm{np},ia}=-F^{(1)}\left(R_a\right)\left(R_i^2\, g_{\mathrm{np,}i} - R_a^2 \, g_{\mathrm{np,}a}\,\right) \cdot 10^{-3}\ .\label{eq:gNPtwo}
\end{equation}
The numerical methods for calculating the np coefficients $g_\mathrm{np}$ are described in Sec.~\ref{sec:NP_num}.

\subsubsection{FS and MS cross term}
Finally, let us turn to the isotope shift contribution which is a mixture of mass shift and field shift. It is convenient to present it as a nuclear-mass-dependent correction to the field-shift constant. The leading nonrelativistic effect comes from the reduced mass and could be included into the isotope shift by the substitution \cite{yerokhin_theory_2019} (see also Ref.~\cite{yerokhin_nonlinear_2020}):
\begin{equation}
    F \rightarrow \mu^3 F = F\left( 1-3\frac{m}{M}+\dots \right) .
\end{equation}
In calculating the cross term below, we use $F^{(1)}(R_a)$ in place of the total $F$, since the difference between the two cases is minuscule in the already small cross term:
\begin{equation}
    \Delta F=-3\frac{m}{M}F^{(1)}(R_a)\ .
\end{equation}
Additionally, the contributions of the same order induced by the specific mass-shift operator are expected to be smaller than the reduced-mass effect and are likewise neglected.

\subsection{General isotope shift formula}\label{sec:all_contrib}
The total isotope shift of an energy level consists of mass shift (Sec.~\ref{sec:MS}), field shift (Sec.~\ref{sec:FS}), and additional terms (Sec.~\ref{sec:AT}). Let us denote $F^{(1)}\equiv F^{(1)}(R_a)$ and $F^{(2)}\equiv F^{(2)}(R_a)$; written as a single sum, the main contributions to the isotope shift between the isotopes $i$ and $a$ are:
\begin{gather}
\Delta E_{ia} = K^{(1)}\left(\frac{m}{M_i}-\frac{m}{M_a}\right)+K^{(2)}\left(\frac{m^2}{M_i^2}-\frac{m^2}{M_a^2}\right)\nonumber\\
+\left[1-f_\mathrm{ho}\left(R_a,R_i\right)\cdot 10^{-3}-3\left(\frac{m}{M_i} -\frac{m}{M_a}\right) \right]F^{(1)} \delta R_{ia}^2 \nonumber\\
-F^{(1)}\left( R_i^2\, g_{\mathrm{np,}i} - R_a^2\, g_{\mathrm{np,}a}\right) \cdot 10^{-3} + F^{(2)}\left(\delta R_{ia}^2\right)^2.\label{eq:IS_TOTAL}
\end{gather}

\section{Many-body perturbation theory}\label{sec:mbpt}
We describe an atom with the relativistic no-pair Dirac-Coulomb Hamiltonian $H$ which is a sum of the zeroth-order Hamiltonian
\begin{equation}
   H_0  = \ \sum_i \Big[ \balpha_i\cdot\bfp_i + \beta_i\,m + V_{\rm nuc}(r_i)
   + U(r_i)\Big]\,,
\end{equation}
where $\bm{\alpha}_i$ and $\beta$ are the Dirac matrices, $V_{\rm nuc}$ is the nuclear Coulomb potential and the residual electron-electron interaction
\begin{equation}
    V_I = \  \sum_{i < j} \Lambda_{++}\, I(r_{ij})\, \Lambda_{++}
  - \sum_{i} \Lambda_+ \, U(r_i)\, \Lambda_+\ .
\end{equation}
In the latter expression, $\Lambda_{++}$ and $\Lambda_{+}$ are the
projection operators to the positive-energy part of the Dirac spectrum, $I$ is the electron-electron interaction operator in the Breit approximation, given by
\begin{equation}
    I(r_{ij}) = \  \frac{\alpha}{r_{ij}} -\frac{\alpha}{2r_{ij}}
   \Big[ \balpha_i\cdot\balpha_j + (\balpha_i\cdot\hat{\bfr}_{ij})(\balpha_j\cdot\hat{\bfr}_{ij}) \Big]\, ,
   \label{eq:3}
\end{equation}
where the first and the second term correspond to the Coulomb and the Breit interaction, respectively. Here $\hat{{\bm r}} = {\bm r}/|\bm{r}|$ and $U$ is a screening potential introduced in the zeroth-order Hamiltonian to partially account for
the electron-electron interaction. An important instance of such a potential is the Dirac-Hartree-Fock potential $V_{\rm HF}$, whose matrix elements are given by
\begin{align}
\lbr i | V_{\rm HF}| j\rbr \equiv ( V_{\rm HF} )_{ij} = \sum_a I_{ai;aj} \,,
\end{align}
where $I_{ab;cd} \equiv I_{abcd} - I_{abdc}$, $I_{abcd} \equiv \lbr ab|I|cd\rbr$, and $I$ is the
operator of the electron-electron interaction defined in Eq.~\eqref{eq:3}. Here, we will adopt the standard notation from Ref.~\cite{blundell_formulas_1987}: the letters $a$, $b$, $c$, $\ldots$ designate occupied core orbitals;
$n$, $m$, $r$, $\ldots$ signify excited orbitals outside the core, including the
valence orbital; $i$, $j$, $k$, $\ldots$ can be either excited or occupied orbitals. The
letter $v$ stands for the valence orbital.

Within the many-body perturbation theory (MBPT), the energy of the valence state $E_v$ is
presented as a perturbation expansion $E_v = E^{(0)}+ E^{(1)} + E^{(2)} + \ldots$ When only the contributions to the valence ionization energy (`val') are considered, the expansion terms can be obtained~\cite{blundell_formulas_1987}:
\begin{subequations}
\begin{align} \label{eq:20}
E^{(0)}_{\rm val} =& \ \vare_v\ ,\\
E^{(1)}_{\rm val} =& \big(V_{\rm HF}-U\big)_{vv}\,,\\
E^{(2)}_{\rm val} =&\  \sum_{amn} \frac{I_{vamn}\, I_{mn;va}}{\epsilon_{av}-\epsilon_{mn}}
 - \sum_{abm} \frac{I_{abmv}\, I_{mv;ab}}{\epsilon_{ab}-\epsilon_{vm}}
 \nonumber \\
 +& 2\,\sum_{am} \frac{(V_{\rm HF}-U)_{am}\, I_{mv;av}}{\vare_a-\vare_m}
 \nonumber \\
 +& \sum_{i\neq v} \frac{(V_{\rm HF}-U)_{vi}\, (V_{\rm HF}-U)_{iv}}{\vare_v-\vare_i}\,,
 \label{eq:21}
\end{align}
\end{subequations}
where $\epsilon_{ab} \equiv \vare_a + \vare_b$. The expressions for the third-order MBPT correction
$E^{(3)}$ are quite lengthy; they are presented in Ref.~\cite{blundell_formulas_1987} and the angular
reduction of these formulae is described in Ref.~\cite{johnson_many-body_1988}. In practical calculations,
it is typical to choose the screening potential $U$ to be the frozen-core Dirac-Fock potential. In
such case, all terms which include the matrix elements of $(V_{\rm HF}-U)$ can be omitted. However, these terms should be preserved if we plan to perturb the above formulae with an additional potential; a perturbation of this kind will be described at the end of this section.

In the present work, we aim to calculate the matrix elements of one-body (field shift) and
two-body (specific mass shift) operators. Moreover, we need to compute the
second-order iterations of such operators. The simplest way to achieve this is to use the
\textit{finite-difference approach}: the perturbing operators are first added to the Hamiltonian
and then the numerical derivative with respect to the perturbations is evaluated. Specifically, for
calculating the field-shift constants $F^{(1)}$ and $F^{(2)}$, we add the perturbing potential
$V_{\rm FS}$ with an arbitrary pre-factor $\lambda$ to the nuclear potential, i.e., $V_{\rm nuc} \to V_{\rm nuc} + \lambda\,V_{\rm FS}$. Then the valence energies $E_v(\lambda)$ of the modified
Hamiltonian are calculated. Finally, the field-shift constants are obtained by computing
numerical derivatives with respect to the parameter $\lambda$:
\begin{equation}
F^{(1)} = \left. \frac{\partial E_v(\lambda)}{\partial \lambda}\right|_{\lambda = 0}\,, \ \ \
F^{(2)} = \left. \frac12 \frac{\partial^2 E_v(\lambda)}{\partial \lambda^2}\right|_{\lambda = 0}\,.
\end{equation}
Similarly, the specific mass-shift constants $K^{(1)}_{\rm SMS}$ and $K^{(2)}_{\rm SMS}$ are
calculated by adding the perturbing potential $V_{\rm SMS}$ to the electron-electron interaction,
$I(r_{ij}) \to I(r_{ij}) + \beta\,V_{\rm SMS}(r_{ij})$ and finding the derivative of the perturbed valence energies $E_v(\beta)$ with respect to $\beta$,
\begin{align}
K^{(1)} = \left. \frac{\partial E_v(\beta)}{\partial \beta}\right|_{\beta = 0}\,, \ \ \
K^{(2)} = \left. \frac12 \frac{\partial^2 E_v(\beta)}{\partial \beta^2}\right|_{\beta = 0}
\,.
\end{align}
The parameters $\lambda$ and $\beta$ are chosen in such a way that the resulting changes in energy are much larger than the round-off errors in the numerical calculation and, at the same time, sufficiently small for the numerical derivative to be stable against variations of $\lambda$ and $\beta$.

\begin{table*}[htb]
\caption{Mass shift and field shift constants in Ca$^+$, see Eq.~\eqref{eq:IS_TOTAL} and Eqs.~(\ref{eq:MS_K1}--\ref{eq:MS_K2}); the respective units are described in Sec.~\ref{sec:units}, Eqs.~(\ref{eq:units_1}--\ref{eq:units_4}). Our results for $\tilde{K}^{(1)}_\mathrm{SMS}$ and $\tilde{F}^{(1)}(R_a)$ are compared with the SMS and FS constants calculated in Refs.~\cite{safronova_third-order_2001,dorne_relativistic_2021}. The dimensionless nuclear charge radius $R_a$ [see Eq.~\eqref{eq:R_def}] belongs to the $^{40}$Ca isotope. The overall sign of the field shift constants from Refs.~\cite{safronova_third-order_2001,dorne_relativistic_2021} was reversed to conform to the definition used in the present work.
\label{tab:const}
}
\begin{ruledtabular}
\begin{tabular}{lccccccc}
&  Units
 &
\multicolumn{1}{l}{}  & \multicolumn{1}{c}{$4s$}
    & \multicolumn{1}{c}{$4p_{1/2}$}
            & \multicolumn{1}{c}{$4p_{3/2}$}
                & \multicolumn{1}{c}{$3d_{3/2}$}
                    & \multicolumn{1}{c}{$3d_{5/2}$}
  \\
\hline\\[-5pt]
$\tilde{K}^{(1)}$  & GHz$\times$amu&
                              & 1324 & 940 & 951 & $-$1136  & $-$1124 \\[5pt]

$\tilde{K}^{(1)}_{\rm SMS}$  & GHz$\times$amu&
                              & $-$251 & $-$221  & $-$206  & $-$2487  & $-$2474 \\
&& MBPT+RPA~\cite{safronova_third-order_2001}
                              & $-$259 & $-$204  & $-$200  & $-$2601  & $-$2595 \\
&& RCCSD(T)~\cite{dorne_relativistic_2021}
                              & $-$243 & $-$208  & $-$204  & $-$2364  & $-$2357 \\[5pt]
$\tilde{K}^{(2)}$ &  GHz$\times$amu$^2$&
                              &$-$3.26 & $-$8.79 & $-$8.78 & $-$3.87 & $-$3.85 \\[5pt]
$\tilde{K}^{(2)}_{\rm SMS}$  & GHz$\times$amu$^2$&
                              &$-$2.53 & $-$8.27 & $-$8.26 & $-$4.49 & $-$4.46
                              \\[8pt]

$\tilde{F}^{(1)}(R_a)$ &  MHz$\times$fm$^{-2}$ & &  266.8 & $-$19.6 & $-$19.9 & $-$112.2  & $-$111.6 \\
&& MBPT+RPA~\cite{safronova_third-order_2001}
                                   &  266.6 & $-$19.6 & $-$19.9 & $-$111.8  & $-$111.2 \\
&& RCCSD(T)~\cite{dorne_relativistic_2021}
                                   &  263.3 & $-$15.9 & $-$19.5 &  $-$93.9  & $-$112.3 \\ [5pt]
$\tilde{F}^{(2)}(R_a)$ & kHz$\times$fm$^{-4}$ & &  $-$89.25 &   6.58 & 6.65 &   37.50  &  37.30 \\[5pt]
%
\end{tabular}
\end{ruledtabular}
\end{table*}

In our calculations, we choose the potential $U$ in the zeroth-order $H_0$ to be the Dirac-Fock potential. When a perturbation is added to the Hamiltonian $H_0$ in the finite-difference approach, one can include this perturbation into the self-consistent procedure of computing the Dirac-Fock potential. It was demonstrated in Ref.~\cite{berengut_isotope-shift_2003} that such an inclusion is advantageous because it accounts for the infinite sequence of diagrams known as the random-phase-approximation (RPA) corrections, thus yielding a substantial improvement in the accuracy of calculations. We adopt this approach and ensure the self-consistency of the Dirac-Fock potential {\em after} the perturbation is added to $H_0$, for each value of the parameters $\lambda$ and $\beta$.

Apart from the finite-difference method, we also implement a more traditional perturbative
approach for calculating the first-order matrix elements of one-body operators---specifically,
the nuclear polarization correction and the higher-order finite nuclear size correction.
Let $V$ be the one-body potential; we consider the linear-in-$V$ perturbations of the MBPT formulas for energy levels. Accordingly, the perturbations of single-electron energies and wave functions are
\begin{align}\label{eq:24}
\vare_i \to \lbr i| V| i\rbr \equiv V_{ii}\,,\ \ \
   |i\rbr \to  |\delta i\rbr = \sum_{k\neq i}\frac{|k\rbr\,V_{ki}}
 {\vare_i-\vare_k}\,.
\end{align}
The perturbations of the first two contributions in Eq.~(\ref{eq:20}) yield
\begin{align}
\label{eq:25}
\delta E^{(0)}_V =&\ V_{vv}\,,\\
\delta E^{(1)}_V =&\ \sum_{k\neq v} \frac{\big(V_{\rm HF}-U\big)_{vk}\, V_{kv}}{\vare_v-\vare_k}
  + 2\, \sum_a I_{\delta\!av;av}\,.
\label{eq:26}
\end{align}
Formulas for the next-order correction $\delta E^{(2)}_V$ are easily obtained from
Eq.~(\ref{eq:21}) by perturbing all single-electron energies and wave functions according to
Eq.~(\ref{eq:24}) and preserving only the part which is linear in $V$.

In order to obtain results equivalent to those delivered by the finite-difference approach, we
need to include the contributions from RPA explicitly. This can be achieved by defining the RPA-corrected single-particle matrix elements as
\cite{safronova_third-order_2001}
\begin{align}
V_{an}^{\,\rm RPA} = V_{an} + \sum_{bm} \frac
 { V_{bm}^{\,\rm RPA}\, I_{am;nb} + I_{ab;nm}\,V_{mb}^{\,\rm RPA}}{\vare_b-\vare_m}\,,
\label{eq:27}
\end{align}
and solving the equation iteratively, obtaining the `dressed' single-particle matrix elements $V_{an}^{\,\rm RPA}$. Then, in the calculation of $\delta E^{(1)}_V$, the resulting $V_{an}^{\,\rm RPA}$ are used instead of $V_{an}$. One can demonstrate that the first-order RPA iteration accounts for a part of $\delta E^{(2)}_V$, while the higher-order RPA iterations correspond to an infinite sequence of higher-order diagrams.

Furthermore, the dressed RPA matrix elements can be implemented in calculations of $\delta E^{(2)}_V$, thus accounting for additional sequence of higher-order contributions. Of course, the double-counting terms should be removed from the RPA-corrected $\delta E^{(1)}_V$ and $\delta E^{(2)}_V$ contributions. A very similar approach was used in Ref.~\cite{safronova_third-order_2001} for the first-order field-shift correction.

\section{Calculation of isotope shift parameters}\label{sec:NUM}
\subsection{Units}\label{sec:units}
Before we consider the specifics of Ca$^+$ numerical calculations, let us briefly discuss the units of isotope shift parameters. According to our definition in Eq.~\eqref{eq:IS_full}, the parameters $K_{ia}$ and $F_{ia}$ have units of energy, since both $m/M_j$ and $R^2_j$ are dimensionless. Such a definition is useful for theoretical treatment, e.g., when we consider the small-parameter expansion in Sec.~\ref{sec:MS}. In the literature, however, the mass shift constant is expressed in GHz$\times$amu and the field shift constant in MHz$\times$fm$^{-2}$. 
In this case, Eq.~\eqref{eq:IS_full} would have the following form
\begin{equation}
  \Delta E_{ia}=\tilde{K}_{ia} \left(\frac{1}{M_i}-\frac{1}{M_a}\right)+\tilde{F}_{ia} \delta\langle r^2\rangle_{ia}\ ,
\end{equation}
The connection between the two definitions of the isotope-shift constants is as follows:
\begin{subequations}
\begin{align}
h\, \tilde{K}^{(1)} &= m K^{(1)}\ ,\label{eq:units_1}\\
h\, \tilde{K}^{(2)} &= m^2 K^{(2)}\ ,\\
h\, \tilde{F}^{(1)} &= \lambdabar_c^{-2} F^{(1)}\ , \\
h\, \tilde{F}^{(2)} &= \lambdabar_c^{-4} F^{(2)}\ .\label{eq:units_4}
\end{align}
\end{subequations}
For clarity, here we explicitly included the Planck constant $h$. In order to make our results more accessible for calculating isotope shifts, in Table~\ref{tab:const} we present the numerical results in the form of $\tilde{K}^{(1)}$, $\tilde{K}^{(2)}$, $\tilde{F}^{(1)}$ and $\tilde{F}^{(2)}$.

It should be mentioned that, in all calculations reported in this work, we do not include the contributions induced solely by the core electrons: they are the same for all states investigated here and cancel when transitions between two states are considered.

\begin{table*}[htb]
\caption{Higher-order field-shift and nuclear-polarization contributions for the ground and first excited states of Ca$^+$ evaluated with different methods. The methods (i), (ii), and (iii) are described in Sec.~\ref{sec:HO_num} and \ref{sec:NP_num}. Higher-order field-shift constant correction $\delta_{R_{46}} F^{(1)}=\delta_{R_{46}} F^{(1)}(R_{40})$ is evaluated for the nuclear charge radii $r_{40} = 3.4776$~fm and $r_{46} = 3.4953$~fm \cite{angeli_table_2013}; note that $R_j=r_j/\lambdabar_C$. Nuclear polarization correction $\Delta E_{\rm np}$ to the electronic energy levels is found for the $^{40}$Ca$^+$ isotope. Units for $\delta_{R_{46}} F^{(1)}$ are kHz$\times$fm$^{-2}$, while $\Delta E_{\rm np}$ is given in MHz.
\label{tab:ho-np}}
\begin{ruledtabular}
\begin{tabular}{lcccccccccc}
\multicolumn{1}{l}{}
 & \multicolumn{2}{c}{$4s$}
    & \multicolumn{2}{c}{$4p_{1/2}$}
            & \multicolumn{2}{c}{$4p_{3/2}$}
                & \multicolumn{2}{c}{$3d_{3/2}$}
                    & \multicolumn{2}{c}{$3d_{5/2}$}
  \\
  \multicolumn{1}{l}{Method} &
  \multicolumn{1}{c}{$\delta_{R_{46}} F^{(1)}$} & \multicolumn{1}{c}{$f_{\rm ho}$} &
  \multicolumn{1}{c}{$\delta_{R_{46}} F^{(1)}$} & \multicolumn{1}{c}{$f_{\rm ho}$} &
  \multicolumn{1}{c}{$\delta_{R_{46}} F^{(1)}$} & \multicolumn{1}{c}{$f_{\rm ho}$} &
  \multicolumn{1}{c}{$\delta_{R_{46}} F^{(1)}$} & \multicolumn{1}{c}{$f_{\rm ho}$} &
  \multicolumn{1}{c}{$\delta_{R_{46}} F^{(1)}$} & \multicolumn{1}{c}{$f_{\rm ho}$}
\\ \hline\\[-5pt]
(i) & -4.810 & 0.0238   & 0.642 & 0.0238   & 0.645 & 0.0237   & 3.814 & 0.0238   & 3.794 & 0.0238 \\
(ii) & -5.354 & 0.0238   & 0.452 & 0.0238   & 0.456 & 0.0237   & 2.011 & 0.0238   & 1.998 & 0.0238 \\
(iii) & -6.202 & 0.0238   & 0.447 & 0.0238   & 0.453 & 0.0237   & 2.519 & 0.0238   & 2.506 & 0.0238 \\
\hline

  &
  \multicolumn{1}{c}{$\Delta E_{\rm np}$} & \multicolumn{1}{c}{$g_{\rm np}$} &
  \multicolumn{1}{c}{$\Delta E_{\rm np}$} & \multicolumn{1}{c}{$g_{\rm np}$} &
  \multicolumn{1}{c}{$\Delta E_{\rm np}$} & \multicolumn{1}{c}{$g_{\rm np}$} &
  \multicolumn{1}{c}{$\Delta E_{\rm np}$} & \multicolumn{1}{c}{$g_{\rm np}$} &
  \multicolumn{1}{c}{$\Delta E_{\rm np}$} & \multicolumn{1}{c}{$g_{\rm np}$}
\\ \hline\\[-5pt]
(i)  & -0.891 & 0.364  & 0.119 & 0.365  & 0.119 & 0.363  & 0.709 & 0.366  & 0.705 &  0.366 \\
(ii) & -0.991 & 0.364  & 0.084 & 0.365  & 0.084 & 0.363  & 0.374 & 0.366  & 0.372 &  0.366 \\
(iii)  & -1.148 & 0.364  & 0.083 & 0.365  & 0.084 & 0.363  & 0.469 & 0.366  & 0.466 &  0.366
\end{tabular}
\end{ruledtabular}
\end{table*}

\begin{table}[tb]
\caption{The higher-order field shift parameter $f_\mathrm{ho}\left(R_a,R_i\right)$ [see Eq.~\eqref{eq:dF_ho}, $A_a=40$] for atomic states of Ca$^+$.
\label{tab:ho}}
\begin{ruledtabular}
\begin{tabular}{cccccc}
  $A_i$& \multicolumn{5}{c}{$f_\mathrm{ho}\left(R_a,R_i\right)$} \\
   & \multicolumn{1}{c}{$4s$} &
                     \multicolumn{1}{c}{$4p_{1/2}$} &
                     \multicolumn{1}{c}{$4p_{3/2}$} &
                     \multicolumn{1}{c}{$3d_{3/2}$} &
                     \multicolumn{1}{c}{$3d_{5/2}$} \\
\hline\\[-7pt]
   42 &   0.0410  &  0.0411  &  0.0410  &  0.0410  &  0.0410  \\
   44 &   0.0542  &  0.0543  &  0.0541  &  0.0541  &  0.0541  \\
   46 &   0.0238  &  0.0238  &  0.0238  &  0.0238  &  0.0238  \\
   48 & $-$0.0007 & $-$0.0007 &  $-$0.0007 &  $-$0.0007 &  $-$0.0007
\end{tabular}
\end{ruledtabular}
\end{table}

\begin{table*}[hbt]
\caption{Nuclear parameters and the nuclear polarization correction coefficient
$g_{\rm np}$ [see Eqs.~(\ref{eq:gNPone}--\ref{eq:gNPtwo})] for various isotopes of Ca$^+$.
The energy of the first nuclear rotational state is $\omega$, $B(E2){\uparrow}$ is
the corresponding nuclear transition strength, and $\sqrt{\langle r^2\rangle}$ is the root-mean-square nuclear charge radius.
 \label{tab:np}}
\begin{ruledtabular}
\begin{tabular}{ccccccccc}
 $A$ & \multicolumn{1}{c}{$\omega$ \cite{raman_transition_2001}}
   & \multicolumn{1}{c}{$B(E2){\uparrow}$ \cite{raman_transition_2001}}
   & \multicolumn{1}{c}{$\sqrt{\langle r^2\rangle}$ \cite{angeli_table_2013}}
   & \multicolumn{5}{c}{$g_{\rm np}$ }
\\

& \multicolumn{1}{c}{[MeV]}
   & \multicolumn{1}{c}{$[e^2{\rm b}^2]$}
   & \multicolumn{1}{c}{[fm]}
       &\multicolumn{1}{c}{$4s$} &
                     \multicolumn{1}{c}{$4p_{1/2}$} &
                     \multicolumn{1}{c}{$4p_{3/2}$} &
                     \multicolumn{1}{c}{$3d_{3/2}$} &
                     \multicolumn{1}{c}{$3d_{5/2}$} \\
\hline\\[-7pt]
  40 &   3.904 &  0.010 &  3.4776 &   0.3643 &    0.3646 &  0.3628 &   0.3658 & 0.3660 \\
  42 &   1.525 &  0.042 &  3.5081 &   0.3961 &    0.3965 &  0.3945 &   0.3977 & 0.3979 \\
  44 &   1.157 &  0.047 &  3.5179 &   0.4122 &    0.4126 &  0.4106 &   0.4139 & 0.4141 \\
  46 &   1.346 &  0.018 &  3.4953 &   0.4120 &    0.4123 &  0.4102 &   0.4138 & 0.4140 \\
  48 &   3.832 &  0.010 &  3.4771 &   0.4242  &   0.4245  & 0.4224 &   0.4262 & 0.4263 \\
\end{tabular}
\end{ruledtabular}
\end{table*}

\subsection{Mass shift}\label{sec:MS_num}
Our calculations of the first- and second-order specific-mass-shift constants $K^{(1)}_{\rm
SMS}$ and $K^{(2)}_{\rm SMS}$ were performed with the finite-difference approach described in Sec.~\ref{sec:mbpt} which reduces the problem at hand to the computation of energies. We included all MBPT corrections to energy up to the third order, i.e., up to the three-photon
exchange. This approach accounts for both the two-photon-exchange corrections to the matrix element
of the SMS operator and an infinite sequence of higher-order corrections delivered by the random-phase approximation. The numerical results for the mass shift constants are presented in Table~\ref{tab:const}. Our values for the first-order constant are in a reasonable agreement with previous calculations \cite{safronova_third-order_2001,dorne_relativistic_2021}.
For the second-order constants, there have been no results reported in the literature.

\subsection{Field shift}\label{sec:FS_num}
The calculations of the first- and second-order field-shift constants $F^{(1)}$ and $F^{(2)}$
were performed with the finite-difference approach as described in Sec.~\ref{sec:mbpt}. We included MBPT corrections to energy up
to the second order, which accounts for all two-photon-exchange
corrections to the matrix element of the field shift operator.

The numerical results for $F^{(1)}$ and $F^{(2)}$ are presented in Table~\ref{tab:const}. 
Our values for the first-order constant $F^{(1)}$ are in a very good agreement with results by
Safronova and Johnson~\cite{safronova_third-order_2001}. 
There is, however, a significant difference
with results of Ref.~\cite{dorne_relativistic_2021}, especially for $4p_{1/2}$ and $3d_{3/2}$
states. 
The probable reason could be that the numerical approach of Ref.~\cite{dorne_relativistic_2021} is
not particularly suitable for computation of the field shift.

\subsection{Additional terms}
\subsubsection{Higher order field shift}\label{sec:HO_num}
In the present work, we consider three different approaches for evaluating the field-shift correction $\delta_{R_i}F^{(1)}\left(R_a\right)$ [see Eq.~\eqref{eq:dF1_ho}]. In the order of the improvement of approximation, they are: (i) Dirac-Fock including core relaxation, based on Eqs.~\eqref{eq:25}~and~\eqref{eq:26}; (ii) random phase approximation (RPA) which relies on adding Eq.~\eqref{eq:27} to the method (i); and, finally, (iii) MBPT2+RPA, in which we added two-photon exchange correction to the method (ii).
Table~\ref{tab:ho-np} compares the results for $\delta_{R_{46}}F^{(1)}(R_{40})$ in Ca$^+$ obtained with different methods. In this table, together with the $\delta_{R_{46}}F^{(1)}(R_{40})$ correction, we display the factor $f_{\rm ho}$ which was introduced in Eq.~\eqref{eq:dF_ho}. We conclude that, to the level of $10^{-3}$ relative accuracy, the factor $f_{\rm ho}$ is both method-independent and the same for all states considered. Our final results for the higher-order field-shift correction are summarized in Table~\ref{tab:ho}. One can note that the results obtained are very close to the hydrogenic $1s$ values. For instance, the hydrogenic result for $f_{ho}(R_{46}, R_{40})$ calculated for the $2s$ state and $Z = 20$ is  $0.02378$, which coincides with the values presented in Table~\ref{tab:ho}.

\subsubsection{Nuclear polarization}\label{sec:NP_num}
The nuclear polarization correction is induced by the operator $V_{\rm np}$ defined in Eq.~\eqref{eq:Vnp-Def}. The single-electron matrix elements of $V_{\rm np}$ were calculated in the same way as in our previous work \cite{muller_nonlinearities_2021}. Specifically, we included the dominant $E2$ nuclear rotational transition and the giant resonance transitions with $L = 0,1,2,3$. The nuclear charge distribution was represented by the two-parameter Fermi distribution model, with the root-mean-square nuclear charge radii taken from Ref.~\cite{angeli_table_2013}. The summation over the Dirac spectrum was performed using the finite-basis-set $B$-spline method \cite{shabaev_dual_2004}. Actual calculations of the one- and two-photon exchange corrections were performed with $N = 50$-$60$ splines of the order 6 and the cavity radius of $35$ a.u. The three-photon exchange corrections were computed with $N = 40$ splines and the partial-wave expansion extended up to $l = 6$. For hydrogenic matrix elements we reproduce the results of Ref.~\cite{nefiodov_nuclear_1996}. A similar calculation has been recently presented in Ref.~\cite{flambaum_nuclear_2021};
the difference is that approximate empirical formulas for $B(E2)$ were used and only the dominant $L = 1$ giant resonance was included.
Based on Eq.~\eqref{eq:gNPone}, the np correction is conveniently expressed in terms of a $g_{\rm np}$ coefficient which is defined by the ratio of nuclear-polarization and field-shift contributions. While calculating this ratio, it is important to use the same method for both contributions. In such case, the ratio will not depend on the method of accounting for electron correlations.

The electron-structure corrections to the np effect are calculated with the same approaches as the higher-order field shift in Sec.~\ref{sec:HO_num}: (i) Dirac-Fock with core relaxation, (ii) RPA, and (iii) MBPT2$+$RPA. The numerical results for both quantities $\Delta E_\mathrm{np}$ and $g_{\rm np}$ delivered by these three methods for the singly-ionized calcium ion $^{40}$Ca$^+$ are listed in Table~\ref{tab:ho-np}. We find that the numerical results expressed in terms of $g_{\rm np}$ do not depend on the method of calculation and only weakly depend on the electronic valence state. This is not surprising since
it is known that for light atoms both the nuclear polarization and the finite nuclear size corrections are roughly proportional to the expectation value of the Dirac $\delta$-function \cite{puchalski_isotope_2006}.

We also find that the results for $g_{\rm np}$ in $^{40}$Ca$^+$ from Table~\ref{tab:ho-np} are very close to the hydrogenic $1s$ value $g_{\rm np}(1s) = 0.360$. The deviation is within 2\%, which is much smaller than the uncertainty associated with the approximate treatment of the nuclear polarization effect. Therefore, for many practical purposes, it is sufficient just to use the hydrogenic values of $g_{\rm np}$.

In Table~\ref{tab:np} we present our results for the nuclear polarization contribution in various isotopes of Ca$^+$. The experimental values of the nuclear quadrupole transition strengths $B(E2)$ and the excitation energies $\omega_L$ originate from Ref.~\cite{raman_transition_2001}, while nuclear charge radii are from Ref.~\cite{angeli_table_2013}.

\subsection{Numerical results}\label{sec:num_uncert}
Our numerical results for the isotope shifts of Ca$^+$ isotopes with $A = 42$, 44, 46, and 48, relative to
the reference isotope $A_0 = 40$ are summarized in Table~\ref{tab:IS_total}.
Note that the numerical results listed in the table do not include
a contribution of the electron core that is the same for all states considered
and cancels for the transition energies.

The uncertainty of our theoretical values of the mass-shift and field-shift constants
stems predominantly from the higher-order electron-correlation effects, because all
computational errors (basis-set truncation, errors of numerical differentiation) are
small. There is no safe way to estimate the omitted electron-correlation effects. To some 
extent, this can be done by comparing calculations performed by different methods. The
specific mass shift is known to be particularly difficult to calculate reliably for
many-electron atoms. The agreement of results obtained by different method summarized in 
Table~\ref{tab:const} for $K^{(1)}_\mathrm{SMS}$ is on the level of 10\%, which could be
taken as an estimation of the uncertainty. 
Calculations of the field shift can generally be performed to a higher accuracy than for the
mass shift. We assume that the uncertainty of our results for $F^{(1)}$ should be within 5\%.

There are no independent calculations to compare with for the second-order isotope shift constants.
We assume here that their uncertainties should be comparable with those of the first-order
constants $K^{(1)}_\mathrm{SMS}$ and $F^{(1)}$.
Furthermore, there are uncertainties originating from the nuclear
model employed in the calculation of the nuclear polarization effect. 
We assume these uncertainties to be on the level of 10\%, which should be considered as an order-of-magnitude estimate \cite{valuev_microscopic_2022, valuev_private_nodate}.

\begin{table*}
\caption{Isotope-shift contributions in MHz and kHz, according to Eq.~\eqref{eq:IS_TOTAL}, with reference isotope $A_0=40$. Note that we do not include the large contributions of core electrons to isotope shifts which would be the same for every single-electron state and cancel out when the difference is considered. MS and FS stand for mass shift and field shift, respectively; `sec.' indicates `second order' and `h.o.' means `higher order'. Nuclear polarization contribution is denoted by `np', while `cross term' means field- and mass-shift cross term.\label{tab:IS_total}}
\begin{ruledtabular}
\begin{tabular}{llcrrrrr}
$A$ & isotope shift & units & $4s$ & $4p_{1/2}$ & $4p_{3/2}$ & $3d_{3/2}$ & $3d_{5/2}$ \\
\hline\\[-7pt]
$42$   	& total & MHz & $-$1519.6 & $-$1122.9 & $-$1136.5 & 1329.8 & 1315.6 \\
\hline
    & MS	       & MHz	&	$-$1576.5 & $-$1119.2 & $-$1132.8 & 1353.4 & 1339.1 \\
	& MS, sec.    & MHz	&	0.190 & 0.511 & 0.511 & 0.225 & 0.224 \\
	& FS		  & MHz	&	56.8 & $-$4.18 & $-$4.24 & $-$23.9 & $-$23.8 \\
	& FS, h.o.	  & kHz	&	$-$2.33 & 0.172 & 0.174 & 0.981 & 0.975 \\
	& FS, sec.	  & kHz	&	$-$4.05 & 0.299 & 0.302 & 1.70 & 1.69 \\
	& np		  & kHz	&	$-$125.1 & 9.24 & 9.31 & 52.8 & 52.5 \\
	& cross term	  & kHz	&	0.111 & $-$0.008 & $-$0.008 & $-$0.047 & $-$0.047 \\
\hline\\[-7pt]
$44$	& total & MHz &  $-$2935.1 & $-$2141.7 & $-$2167.7 & 2553.2 & 2526.0 \\
\hline
        & MS		& MHz	& $-$3010.4 & $-$2137.1 & $-$2163.1 & 2584.3 & 2557.0 \\
		& MS, sec.	& MHz	& 0.354 & 0.955 & 0.954 & 0.420 & 0.418 \\
		& FS		& MHz	& 75.2 & $-$5.54 & $-$5.62 & $-$31.6 & $-$31.5 \\
		& FS, h.o.	& kHz	& $-$4.08 & 0.301 & 0.304 & 1.71 & 1.70 \\
		& FS, sec.	& kHz	& $-$7.09 & 0.523 & 0.529 & 2.98 & 2.96 \\ 
		& np		& kHz	& $-$185.5 & 13.7 & 13.8 & 78.4 & 78.0 \\
		& cross term	& kHz	& 0.282 & $-$0.021 & $-$0.021 & $-$0.118 & $-$0.118 \\
\hline\\[-7pt]

$46$	& total & MHz &  $-$4287.2 & $-$3068.2 & $-$3105.5 & 3695.8 & 3656.6 \\
\hline        
        & MS		& MHz	& $-$4320.5 & $-$3067.1 & $-$3104.4 & 3709.0 & 3669.7 \\
		& MS, sec.	& MHz	& 0.498 & 1.34 & 1.34 & 0.590 & 0.587 \\
		& FS		& MHz	& 32.9 & $-$2.42 & $-$2.46 & $-$13.9 & $-$13.8 \\
		& FS, h.o.	& kHz	& $-$0.784 & 0.058 & 0.059 & 0.330 & 0.328 \\
		& FS, sec.	& kHz	& $-$1.36 & 0.100 & 0.101 & 0.571 & 0.568 \\
		& np		& kHz	& $-$167.5 & 12.3 & 12.4 & 70.9 & 70.5 \\
		& cross term	& kHz	& 0.177 & $-$0.013 & $-$0.013 & $-$0.074 & $-$0.074 \\
\hline\\[-7pt]

$48$	& total & MHz & $-$5528.5 & $-$3922.6 & $-$3970.3 & 4746.9 & 4696.6 \\
\hline        
        & MS		& MHz	& $-$5528.0 & $-$3924.4 & $-$3972.1 & 4745.6 & 4695.4 \\
		& MS, sec.	& MHz	& 0.624 & 1.68 & 1.68 & 0.741 & 0.737 \\
		& FS		& MHz	& $-$0.928 & 0.068 & 0.069 & 0.390 & 0.388 \\
		& FS, h.o.	& kHz	& $-$0.00065 & 0.00005 & 0.00005 & 0.00027 & 0.00027 \\
		& FS, sec.	& kHz	& $-$0.0011 & 0.0001 & 0.0001 & 0.0005 & 0.0005 \\
		& np		& kHz	& $-$192.9 & 14.2 & 14.3 & 81.8 & 81.2 \\
		& cross term & kHz	& $-$0.0064 & 0.0005 & 0.0005 & 0.0027 & 0.0027 \\

\end{tabular}
\end{ruledtabular}
\end{table*}

\section{King plots}\label{sec:KP}
\subsection{Theoretical introduction}
\begin{center}
		\begin{figure}[bb]
			\centering
			\includegraphics[width=\linewidth]{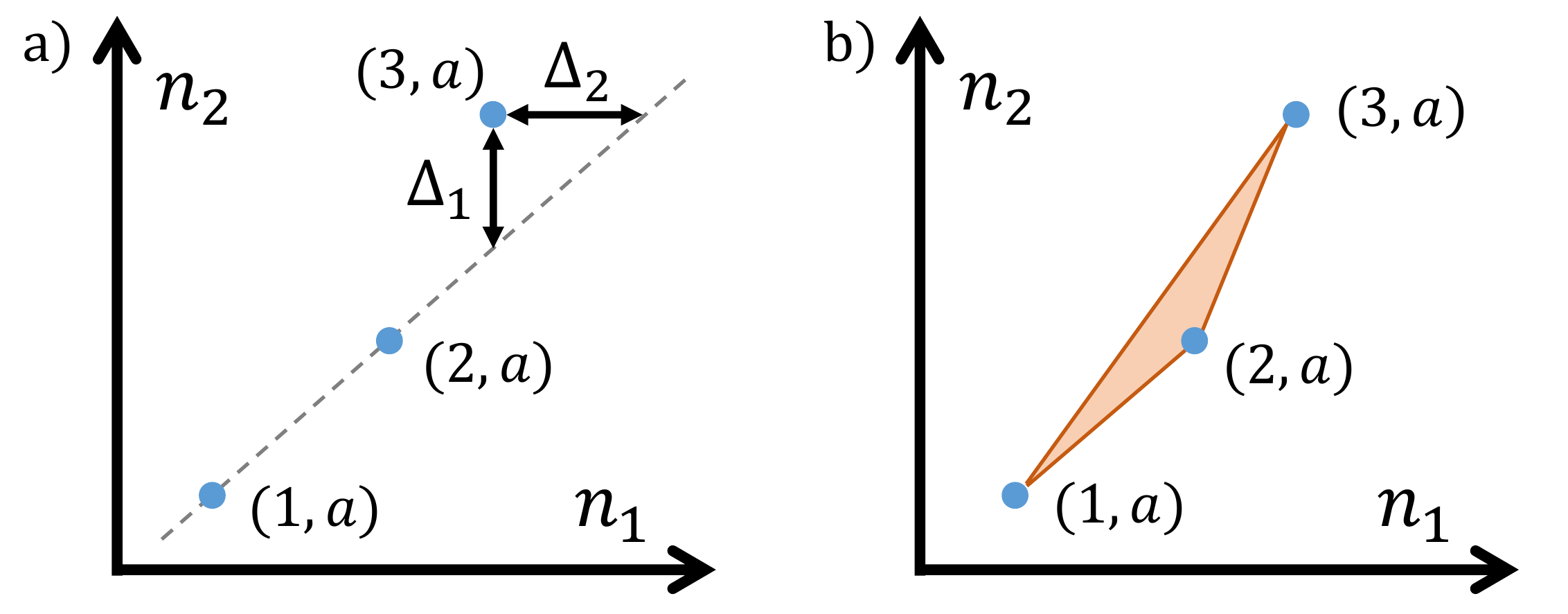}
			\caption{Two ways to evaluate the nonlinearity of a King plot. Here $n_{1,2}$ are modified isotope shift frequencies and $(i,a)$ indicates the isotope pair which corresponds to the given point; $i=1,2,3$ and we assume $a$ to be the reference isotope $^{40}$Ca. Method \textbf{a)} characterizes the average difference $\Delta_\mathrm{line}=\left(\Delta_1+\Delta_2\right)/2$ between the outlier point and the line defined by the two other points. Method \textbf{b)} quantifies the area of the triangle defined by the three points, see Eq.~\eqref{eq:NL_triangle} and Ref.~\cite{berengut_probing_2018}.}
			\label{fig:KP_nonlin}
		\end{figure}
\end{center}

The method of King plots \cite{king_isotope_2013} is a popular way to analyze the experimentally measured isotope shifts. There are different ways to construct King plots from spectroscopy data to serve similar purposes, see Refs.~\cite{king_isotope_2013, counts_evidence_2020}; we will consider the most widely used version. The core idea of a King plot is that, to the leading order, isotope shifts depend linearly on $m/M_i$ and $R^2_i$, see Eq.~\eqref{eq:IS_1st}. Hence, if one considers isotope shifts of two different transitions, it is possible to eliminate the poorly known $R^2_i$ from a system of linear equations. To illustrate this approach, let $\Delta \nu=\Delta E^{(e)}-\Delta E^{(g)}$ be a transition frequency between an excited and ground atomic energy levels. Then, to the first order
\begin{align}
\begin{cases}
\Delta \nu_{1,ia}=K_{\nu_1} \mu_{ia}+F_{\nu_1} \delta R^2_{ia}\ ,\\
\Delta \nu_{2,ia}=K_{\nu_2} \mu_{ia}+F_{\nu_2} \delta R^2_{ia}\ ,
\end{cases}
\end{align}
with $\mu_{ia}=\left(m/M_i-m/M_a\right)$. Consequently, we can write a linear relation
\begin{equation}
    n_{2,ia}=\frac{F_{\nu_2}}{F_{\nu_1}}\ n_{1,ia}+\left(K_{\nu_2}-\frac{F_{\nu_2}}{F_{\nu_2}}K_{\nu_1}\right)\ ,\label{eq:KP_linear}
\end{equation}
where $n_{k,ia}=\Delta \nu_{k,ia}/\mu_{ia}$ are modified frequencies ($k=1,2$). The plot of $n_{2,ia}$ against $n_{1,ia}$ for different isotope pairs $(i,a)$ is called King plot (see Fig.~\ref{fig:KP_nonlin}) and, to the first order in isotope shift, it is linear.

The higher-order corrections described in Sec.~\ref{sec:IS_overall} and summarized in Eq.~\eqref{eq:IS_TOTAL} distort the linear relation \eqref{eq:KP_linear}. To quantify this deviation from linearity, we will apply two different methods shown schematically in Fig.~\ref{fig:KP_nonlin}. The first method relies on averaging the difference between the outlier point and the line defined by two other points:
\begin{equation}\label{eq:NL_line}
    \Delta_\mathrm{line}=(\Delta_1+\Delta_2)/2,\quad \Delta_{1,2}>0\ .
\end{equation}
To scale $\Delta_\mathrm{line}$ back to frequency units, we multiply its value by $|m/M_j-m/M_a|$, where ($j,a$) is the isotope pair of the outlier point.

The second method considers the area of a triangle defined by three points of a King plot \cite{berengut_probing_2018}. To formalize this approach, we need to define `isotope shift vectors' whose coordinates correspond to modified frequency shifts in different isotope pairs:
\begin{equation}\label{eq:IS_vec1}
    \vec{n}_{k}=(n_{k,1a},\ n_{k,2a},\ n_{k,3a}).
\end{equation}
The area of the King-plot triangle can be calculated as \cite{berengut_probing_2018, berengut_generalized_2020}
\begin{equation}\label{eq:NL_triangle}
\Delta_{V_2}=\frac{1}{2}\,\mathrm{det}\left(\vec{n}_{1},\ \vec{n}_{2},\ \vec{\mathbb{1}}_3\right),
\end{equation}
where $\vec{\mathbb{1}}_3\equiv\left(1,1,1\right)$. Again, in order to scale $\Delta_\mathrm{V_2}$ to frequency units, we multiply it by the second power of the mean value $\overline{|m/M_i-m/M_a|}^2$, where ($i,a$) are all four isotope pairs.

The latter approach is easily generalized to include more isotopes and more transitions \cite{berengut_generalized_2020}. For instance, the isotope shifts in Ca$^+$ can be measured in four pairs of isotopes: the reference $A=$40 and each of the $A=$42, 44, 46, 48 isotopes. In this way, we would obtain four-dimensional isotope-shift vectors:
\begin{equation}
    \vec{n}_k=\left(n_{k,1a},\ n_{k,2a},\ n_{k,3a},\ n_{k,4a} \right).\label{eq:IS_vec1}\\
\end{equation}
The nonlinearity is defined by
\begin{equation}
    \Delta_{V_3}=\frac{1}{6}\,\mathrm{det}\left(\vec{n}_{1},\ \vec{n}_{2},\ \vec{n}_{3},\ \vec{\mathbb{1}}_4\right) ,\label{eq:3D_NL}
\end{equation}
where $\vec{\mathbb{1}}_4=\left(1,1,1,1\right)$. We scale $\Delta_\mathrm{V_3}$ back to frequency units by introducing the mean-value factor $\overline{|m/M_i-m/M_a|}^3$, where ($i,a$) are all four isotope pairs. This case corresponds to a three-dimensional King plot. 

In the present work we analyze nonlinearities both for the two- and three-dimentional King plots. In former case, we consider
two sets of isotopes: $(A_i,40)$ with $A_i=$42, 44, 46 and $A_i=$42, 44, 48. Note that the charge radii of $^{40}$Ca and $^{48}$Ca are almost equal, while the radii of the $A=$42, 44, 46 isotopes are larger than both, as can be seen from Table~\ref{tab:np}; this fact may influence the pattern of King plot nonlinearity in a way that is specific for calcium.

\subsection{King-plot nonlinearity in Ca$^+$}
We calculate two-dimensional King-plot nonlinearities (NLs) with two methods illustrated in Fig.~\ref{fig:KP_nonlin}, for two pairs of transitions: ($3d_{3/2}\rightarrow 4s$; $3d_{5/2}\rightarrow 4s$) and ($4p_{1/2}\rightarrow 4s$; $3d_{3/2}\rightarrow 4s$). The first pair consists of two narrow transitions and is suitable for the search of possible new-physics effects \cite{solaro_improved_2020}.
However, this transition pair poses a challenge for theoretical predictions of NLs, as it involves significant cancellations between the isotope shifts. On the other hand, the second pair of transitions does not have such strong cancellations and, consequently, is better suited for detecting King-plot nonlinearities originating from the standard model, as will be discussed below.

Since there are four isotope pairs available for Ca$^+$---four points on a 2D King plot---we calculate NLs both for the `third point' ($A=$42, 44, 46 vs. $A=$40) and the `fourth point' ($A=$42, 44, 48 vs. $A=$40). Additionally, we determine the 3D King plot NL as defined in Eq.~\eqref{eq:3D_NL}, for the following set of transitions: $3d_{3/2}\rightarrow 4s$, $3d_{5/2}\rightarrow 4s$, and $4p_{1/2}\rightarrow 4s$.

Our numerical results are presented in Tables~\ref{tab:NL_2D} and \ref{tab:NL_3D}
for the NL of the 2D and 3D King plots, respectively. Individual contributions to 2D NL are roughly additive; in Table~\ref{tab:NL_2D} we present the total NL results as well as
the individual contributions expressed in percentage of the total values. On the other hand, individual contributions to the 3D NL turn out to be non-additive. So, we successively \textit{exclude} each of the terms and list the resulting change of NL values in Table~\ref{tab:NL_3D}.

Examining individual contributions to NL, we observe that the second-order and higher-order FS corrections and the cross term are essentially negligible for Ca$^+$ and can be omitted in future studies. This was expected, since the nuclear-size effects tend to be small for light atoms. The main source of NL in Ca$^+$ is the second-order mass shift, which is consistent with our previous findings for argon \cite{yerokhin_theory_2019}. Quite surprisingly, however, we find a large contribution from the nuclear polarization for the 4$^\mathrm{th}$-point 
($3d_{3/2},3d_{5/2}\rightarrow 4s$) 2D King-plot and for the 3D King plot. A possible reason is the irregular behavior of the nuclear charge radii: the charge radii of $^{40}$Ca and $^{48}$Ca are almost equal, whereas the radii of $^{42,44,46}$Ca are larger than both of them (see Table~\ref{tab:np}).

One might notice that the methods a) and b) produce NLs that are, first, of different units and,
second, very different numerically. This is not surprising since the method a) measures a linear
distance whereas the method b) measures the area of a triangle. Naturally, both methods describe the same phenomenon and thus the resulting NLs correspond to the same experimental uncertainty which would allow to detect them. We checked this numerically by a Monte Carlo simulation and
confirmed that the experimental errors at which the NLs become visible is the same for a) and b) methods of the NL determination.

We now address the question of how accurately the present theory can predict the NLs of King plots within the standard model. Theoretical accuracy of isotope-shift (particularly, the specific mass shift) constants of many-electron systems achievable in modern calculations is not very high---on the level of few percent. Initially, it was assumed that, because the NLs are very small, even an approximate theoretical result would provide valuable information for searching for new physics with Kings plots. However, as has been shown in our previous work \cite{yerokhin_nonlinear_2020}, the numerical values of NLs in hydrogenlike ions turn out to be very sensitive to the experimental uncertainties of the nuclear charge radii, which places limits on theoretical predictions of NLs. In the present work, we studied the sensitivity of the numerical values of NLs in Ca$^+$ to the theoretical uncertainties of isotope-shift parameters and to the experimental errors of nuclear charge radii. Since the standard methods of error propagation do not work in this case, we used a Monte Carlo simulation. 
%
%
Each of the computed isotope-shift constants was represented by a set of quasirandom normally-distributed numbers, with parameters of the normal distribution defined by the central value and the uncertainty of the corresponding isotope-shift constant. The NL errors were obtained by analyzing the resulting distributions of the NL values. The uncertainty estimates described in Sec.~\ref{sec:num_uncert} were employed for the isotope-shift constants. For the nuclear charge radii, we used the errors from Ref.~\cite{angeli_table_2013}, specifically the systematic uncertainty of 0.0020~fm for all isotopes and the relative uncertainty of 0.0009~fm of nuclear charge radii of isotopes $A = $42, 44, 46, and 48 relative to $A=40$.

The numerical results are summarized in Table~\ref{tab:NL_2D_2}. The two methods of the NL determination yield very similar results, therefore it is sufficient to present the values only for the method a). We find that for the ($3d_{3/2},\, 3d_{5/2}\rightarrow 4s$) pair the theoretical uncertainty $\epsilon_\mathrm{const}$ is an order of magnitude larger than the central value. This is not surprising, given the strong cancellations between the isotope shifts of the two transitions. As is evident from Table~\ref{tab:IS_total}, we cannot reliably predict the isotope shift of the $3d_{5/2}$--$3d_{3/2}$ fine-structure difference, and this fact leads to a large uncertainty in the corresponding NLs. Hence, for the ($3d_{3/2},\, 3d_{5/2}\rightarrow 4s$) pair only an upper limit of NL can be obtained: we conclude that the standard-model NLs for this transition pair in Ca$^+$ might be observed when the experimental accuracy is below 200~Hz. As a matter of fact, such accuracy is already achievable in modern experiments with Ca$^+$ \cite{knollmann_part-per-billion_2019,solaro_improved_2020}, although no confirmed NLs have been reported so far. 

The second pair of transitions considered in this work, ($4p_{1/2}\rightarrow 4s$; $3d_{3/2}\rightarrow 4s$), involves transitions with different principal quantum numbers $n$, hence the cancellation between the isotope shifts is much weaker. Accordingly, the NLs predicted for this KP transition pair are significantly larger and can be seen already at 
the experimental accuracy of about 1.5~kHz. The purely theoretical uncertainty---which stems from the  uncertainty of isotope-shift constants---amounts to  $\sim$20\% of the NL value, which means that, in this case, our theory can quantitatively predict a NL. Additionally, the nuclear-radii uncertainty turns out to be about twice as large as the purely theoretical one. This means that if observed, the 
($4p_{1/2}\rightarrow 4s$; $3d_{3/2}\rightarrow 4s$) King-plot can be used to improve our knowledge of nuclear charge radii.

\begin{table}[tb]
\caption{Two-dimensional King-plot NLs. Methods a) and b) are depicted in Fig.~\ref{fig:KP_nonlin}. The NL units are Hz for method a) and kHz$^2$ for method b). The `3$^\mathrm{rd}$ point' column corresponds to the isotope shifts in $A=42,44,46$ isotopes and the `4$^\mathrm{th}$ point' to the isotope shifts in $A=42,44,48$ isotopes with respect to $A_a=40$. Approximate percentages of each higher-order term contribution to the total NLs are given in respective columns. The abbreviations for the individual terms are the same as in Table~\ref{tab:IS_total}. \label{tab:NL_2D}}
\begin{ruledtabular}
\begin{tabular}{llcccc}
transitions & contribution & \multicolumn{2}{c}{3$^\mathrm{rd}$ point} & \multicolumn{2}{c}{4$^\mathrm{th}$ point} \\
 &   & a) & b)  & a) & b) \\
\hline
$3d_{3/2}\rightarrow 4s;$
                &  total    &  44  &  1.1$\times10^3$ &       180 &  3.3$\times10^3$   \\
$3d_{5/2}\rightarrow 4s.$
                &  MS, sec. &  99.2\%  &  99.2\%  &    24.7\%  &  24.7\%  \\
                &  FS, h.o. &  0.1\%  &  0.1\%  &      0.0\%  &  0.0\%  \\
                &  FS, sec. & 0.6\%  &  0.6\%  &       0.2\%  &  0.2\% \\
                & np        & 0.2\% &  0.2\%  &        75.1\% &  75.0\% \\
                & cross term  & 0.0\% &  0.0\% &       0.0\% &  0.0\% \\
\hline
$4p_{1/2}\rightarrow 4s;$
   &  total    &  1.4$\times10^4$  &  2.8$\times10^5$   & 1.4$\times10^4$  &  2.2$\times10^5$ \\
$ 3d_{3/2}\rightarrow 4s.$
   &  MS, sec. &  99.7\%  &  99.7\%     &   99.5\%  &  99.5\%   \\
   &  FS, h.o. &  0.1\%  &  0.1\%      & 0.1\%  &  0.1\%   \\
   &  FS, sec. &  0.0\%  &  0.0\%     & 0.1\%  &  0.1\%  \\
   &  np        &  0.2\%  & 0.2\%     & 0.3\%   &  0.3\%   \\
   & cross term  & 0.0\%  & 0.0\%     & 0.0\%  & 0.0\%   \\
\end{tabular}
\end{ruledtabular}
\end{table}

\begin{table}[tb]
\caption{The three-dimensional King plot NLs, see Eq.~\eqref{eq:3D_NL}. Contributions to the three-dimensional NL are non-additive; we consecutively \textit{exclude} each term from the total sum and record how the NL value is impacted. The second column names the one term excluded from the sum, whereas the third column shows the \textit{change} of the NL in percentage of the `total' value with all terms included. The abbreviations for the individual terms are the same as in Table~\ref{tab:IS_total}.\label{tab:NL_3D}}
\begin{ruledtabular}
\begin{tabular}{p{1.9cm} p{3.0cm} c}
transitions & contribution & NL[kHz$^3$] \\
\hline
$4p_{1/2}\rightarrow 4s$;    &  total         &   8.2$\times10^3$  \\
$3d_{3/2}\rightarrow 4s$;    &  no MS, sec.   &   99.9\%  \\
$3d_{5/2}\rightarrow 4s$.    &  no FS, h.o.   &   0.0\%  \\
                             &  no FS, sec.    &   0.0\%   \\
                             &  no np          &   99.9\%   \\
                             &  no cross term  &   0.0\%  \\
\end{tabular}
\end{ruledtabular}
\end{table}

\begin{table}[tb]
\caption{NLs and their uncertainties for the two-dimensional King plots. For each entry, the upper line shows a NL calculated with method a) and its uncertainties, while the lower line gives the maximal experimental uncertainties of (both) transition frequencies that allow to detect the NL values listed in the upper line. $\epsilon_\mathrm{const}$ is the theoretical error of the NL value stemming from the numerical uncertainty of isotope-shift constants, whereas $\epsilon_\mathrm{radii}$ is the NL uncertainty originating from the experimental values of nuclear charge radii. The units are kHz.
\label{tab:NL_2D_2}}
\begin{ruledtabular}
\begin{tabular}{llccc}
Transitions &  & \multicolumn{1}{c}{Value}
      & \multicolumn{1}{c}{$\epsilon_\mathrm{const}$} 
      & \multicolumn{1}{c}{$\epsilon_\mathrm{radii}$} \\ 
\hline
$3d_{3/2}\rightarrow 4s;$
                &  3$^{\mathrm{rd}}$ point  &  0.044  & 1.5   & 0.013 \\
$3d_{5/2}\rightarrow 4s.$
                &                           & (0.005) & (0.20) & (0.002) \\
                &  4$^{\mathrm{th}}$ point  &  0.181  & 1.5   & 0.023 \\
                &                           & (0.012) & (0.12) & (0.002) \\
\hline
$4p_{1/2}\rightarrow 4s;$
                &  3$^{\mathrm{rd}}$ point  &  14  & 2.4   & 4.0 \\
$3d_{3/2}\rightarrow 4s.$
                &                           & (1.3) & (0.3) & (0.45) \\
                &  4$^{\mathrm{th}}$ point  &  14  & 2.5   & 7.0 \\
                &                           & (0.8) & (0.17) & (0.4) \\
\end{tabular}
\end{ruledtabular}
\end{table}

\section{Conclusion}\label{sec:conclusion}
We performed a detailed study of isotope shifts in Ca$^+$ transition energies. The calculations of the first-order mass-shift and field-shift isotope-shift constants were carried out within the relativistic many-body perturbation theory for the $4s$, $4p_{1/2,\, 3/2}$, and $4d_{3/2,\, 5/2}$ states of Ca$^+$. The results are in good agreement with the previous calculations which were performed using many-body perturbation theory \cite{safronova_third-order_2001} and coupled-cluster method \cite{dorne_relativistic_2021}. 

For the first time, the higher-order isotope-shift effects responsible for the standard-model (SM) nonlinearities in King plots were calculated: second-order mass shift, second- and higher-order field shifts, and nuclear polarization. We analyzed the resulting Kings-plot nonlinearities in Ca$^+$ with three different methods and demonstrated that the dominant contributions originate from the second-order mass shift and nuclear polarization. 

Two pairs of transitions were examined in this work. The first pair,
($3d_{3/2}\rightarrow 4s$; $3d_{5/2}\rightarrow 4s$), consists of two narrow transitions that can be measured very accurately, which makes it suitable for new-physics searches with King plots. However, for this pair the present theory can only predict an upper limit of the nonlinearities caused by the SM effects, owing to large cancellations between the isotope shits. Based on our results, we conclude that the SM nonlinearities in this transition pair can become visible when the experimental accuracy is below 200~Hz, an accuracy already achieved in modern experiments with Ca$^+$ \cite{knollmann_part-per-billion_2019,knollmann_erratum_2023,solaro_improved_2020}. So, as long as experiments do not detect any King-plot nonlinearity, this fact can be used for placing bounds on possible new-physics effects, see Ref.~\cite{berengut_probing_2018}. However, as soon as the King plot is measured to be nonlinear, it would be hard to discern whether this nonlinearity is due to a SM effect or a new-physics interaction. 

The second transition pair investigated in this work, ($4p_{1/2}\rightarrow 4s$; $3d_{3/2}\rightarrow 4s$), involves two very different transitions. The line profile of the $4p$-$4s$ transition is rather wide and cannot be measured as accurately as the $3d$-$4s$ one.
However, the cancellation between the isotope shifts is relatively small in this case and the theory can provide a quantitative prediction of the SM nonlinearity. As follows from Table~\ref{tab:NL_2D_2}, the
King-plot nonlinearity for this transition pair 
can be observed already at the 1~kHz level of experimental uncertainty, 
which is feasible in modern experiments. 
We conclude that the  ($4p_{1/2}\rightarrow 4s$; $3d_{3/2}\rightarrow 4s$) transition pair is a promising candidate  for the experimental
identification of the SM King-plot nonlinearity and for obtaining information about nuclear charge radii.

\section*{Acknowledgements}
This work was funded by the Deutsche Forschungsgemeinschaft (DFG, German Research Foundation) under Germany’s Excellence Strategy -- EXC-2123 QuantumFrontiers -- 390837967 and under the project No. SU 658/4-2.


\bibliographystyle{apsrev}

\end{document}